\documentclass[]{spie}  %>>> use this instead for A4 paper
\usepackage{amsmath,amsfonts,amssymb}
\usepackage{graphicx}
\usepackage{subfig}
\usepackage[colorlinks=true, allcolors=blue]{hyperref}

\title{Development of the Low Frequency Telescope focal plane detector arrays for LiteBIRD}

\author[a]{Tommaso Ghigna}
\author[b]{Aritoki Suzuki}
\author[c]{Benjamin Westbrook}
\author[c]{Christopher Raum}
\author[a]{Hiroki Akamatsu}
\author[c]{Shawn Beckman}
\author[c]{Nicole Farias}
\author[a]{Tijmen de Haan}
\author[d]{Nils Halverson}
\author[a,e,f,g,h]{Masashi Hazumi}
\author[i]{Johannes Hubmayr}
\author[d]{Greg Jaehnig}
\author[c,b,a]{Adrian T. Lee}
\author[j]{Samantha L. Stever}
\author[a]{Yu Zhou}
\author[ ]{for~the~\textit{LiteBIRD}~Collaboration}
\affil[a]{QUP (WPI), KEK, Tsukuba, Ibaraki 305-0801, Japan}
\affil[b]{LBNL, Berkeley, California 94720, USA}
\affil[c]{UC Berkeley, Berkeley, California 94720, USA}
\affil[d]{CASA, University of Colorado, Boulder, Colorado 80309, USA}
\affil[e]{Kavli IPMU (WPI), UTIAS, The University of Tokyo, Kashiwa, Chiba 277-8583, Japan}
\affil[f]{JAXA, ISAS, Sagamihara, Kanagawa 252-5210, Japan}
\affil[g]{SOKENDAI, Miura District, Kanagawa 240-0115, Hayama, Japan}
\affil[h]{IPNS, KEK, Tsukuba, Ibaraki 305-0801, Japan}
\affil[i]{NIST, Boulder, Colorado 80305, USA}
\affil[j]{Okayama University, Department of Physics, Okayama 700-8530, Japan}

\authorinfo{Further author information: (Send correspondence to T.G.) E-mail: tommaso.ghigna@kek.jp}

\begin{document} 
\maketitle

\begin{abstract}
LiteBIRD, a forthcoming JAXA mission, aims to accurately study the microwave sky within the 40-400 GHz frequency range divided into 15 distinct nominal bands. The primary objective is to constrain the CMB inflationary signal, specifically the primordial B-modes. LiteBIRD targets the CMB B-mode signal on large angular scales, where the primordial inflationary signal is expected to dominate, with the goal of reaching a tensor-to-scalar ratio sensitivity of $\sigma_r\sim0.001$.
LiteBIRD frequency bands will be split among three telescopes, with some overlap between telescopes for better control of systematic effects. Here we report on the development status of the detector arrays for the Low Frequency Telescope (LFT), which spans the 34-161 GHz range, with 12 bands subdivided between four types of trichroic pixels consisting of lenslet-coupled sinuous antennas. The signal from the antenna is bandpass filtered and sensed by AlMn Transition-Edge Sensors (TES). We provide an update on the status of the design and development of LiteBIRD's LFT LF1 (40-60-78 GHz),  LF2 (50-68-89 GHz) pixels. We discuss design choices motivated by LiteBIRD scientific goals. In particular we focus on the details of the optimization of the design parameters of the sinuous antenna, on-chip bandpass filters, cross-under and impedance transformers and all the RF components that define the LF1 and LF2 pixel detection chain. We present this work in the context of the technical challenges and physical constraints imposed by the finite size of the instrument.
\end{abstract}


\section{INTRODUCTION}
\label{sec:intro}  
Studying the primordial CMB (Cosmic Microwave Background) B-mode signal stands out as a primary objective in modern cosmology. While still an unanswered question, existing theories predict a rapid and extensive inflationary expansion of the Universe. The most effective tool at our disposal for investigating inflation is to measure the polarized CMB signal \cite{zaldarriaga_seljak97, kamionkowski97}. When observing large angular scales ($\ell \lesssim 200$), it is expected that the dominant constituent of the polarized B-mode signal is a primordial inflationary component, given a tensor-to-scalar ratio $r\gtrsim0.01$, or even $r\gtrsim0.001$ for $\ell < 10$.

Detecting the primordial B-mode signal and constraining inflation are the main goals of the LiteBIRD space mission \cite{Hazumi2021PTEP} as well as several other ongoing and planned CMB experiments \cite{Suzuki2016, simonsObs2019, cmbs4}. A common feature of most CMB experiments is the use of Transition-Edge Sensor (TES) bolometers to detect the incoming radiation \cite{Irwin2005}. This technology has undergone intense development in the last two decades. 
\begin{figure}[tp]
    \centering
    \includegraphics[width=1\textwidth]{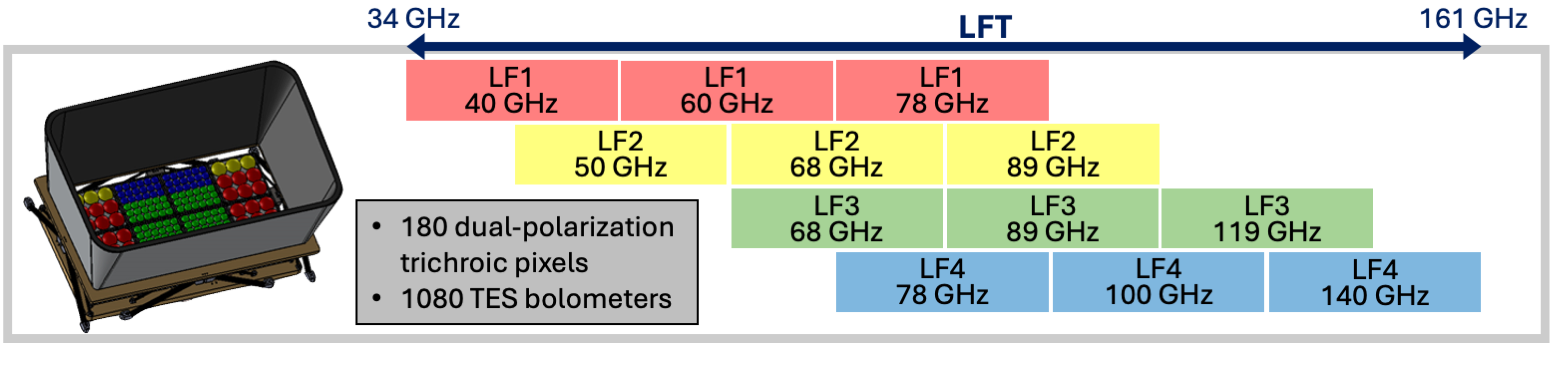}
    \caption{A summary of LFT FPU (focal plane unit). A total of 180 lenslet-coupled sinuous antenna pixels are arranged in a rectangular focal plane divided among 8 FPSAs (focal plane sub-assemblies): 4 LF12 FPSAs (LF1 and LF2 pixels) at the edges of the FPU and 4 LF34\cite{Raum2024} FPSAs (LF3 and LF4 pixels) at the center of the FPU. The CAD rendering of LFT focal plane is adapted from \citenum{Hazumi2021PTEP}.}%
    \label{fig:lftfpu}
\end{figure}
\begin{figure}[bp]
    \centering
    \includegraphics[width=1\textwidth]{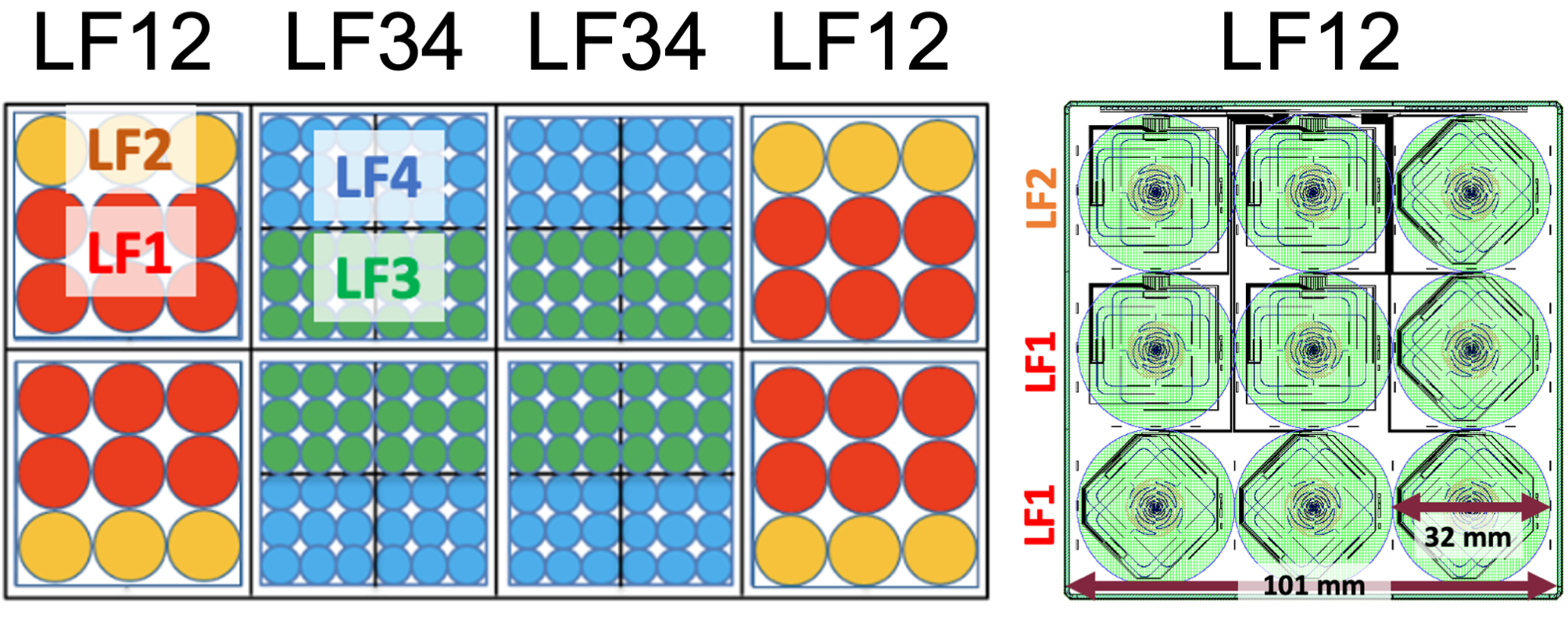}
    \caption{\textbf{Left:} A summary of the LFT FPU layout showing the total number of detector wafers and type (LF12 and LF34). Pixel types are highlighted in red for LF1, yellow for LF2, green for LF3 and blue for LF4. \textbf{Right:} The layout of a LF12 wafer showing 6 LF1 pixels (bottom 2 rows) and 3 LF2 pixels (top row). The LFT focal plane layout on the left is adapted from \citenum{Hazumi2021PTEP}.}%
    \label{fig:FPUandLF12wafer}
\end{figure}

In order to be used for millimetre-wave detection, a common technique in modern experiment consists in coupling TES detectors to an antenna to collect the incoming radiation. At present a handful of technologies are being used for CMB observations. Among them we can find lenslet-coupled dual-polarised sinuous antenna \cite{Suzuki2016, suzuki_phd} and horn-coupled OMTs \cite{Hubmayr2022} (orthomode transducer). LiteBIRD will implement both technologies in three separate telescopes mounted on the satellite payload: Low Frequency Telescope (LFT - 34-161 GHz), Mid Frequency Telescope (MFT - 89-225 GHz) and High Frequency Telescope (HFT - 166-448 GHz)\cite{litebird_ptep}. Sinuous antennas will be used for the lower frequency range (LFT and MFT), while horns antenna will cover the highest frequencies (HFT)\cite{Hubmayr2018}. Aside for the different antenna types, the two technologies share many common features. Detector arrays are microfabricated on silicon wafers and the signal collected by each antenna is routed to the detector by means of microstrip (or coplanar) transmission lines. The signal is filtered by on-chip band defining filters and dissipated on the TES island by a resistor matched to the transmission line.
\begin{figure}[tp]
    \centering
    \includegraphics[width=1\textwidth]{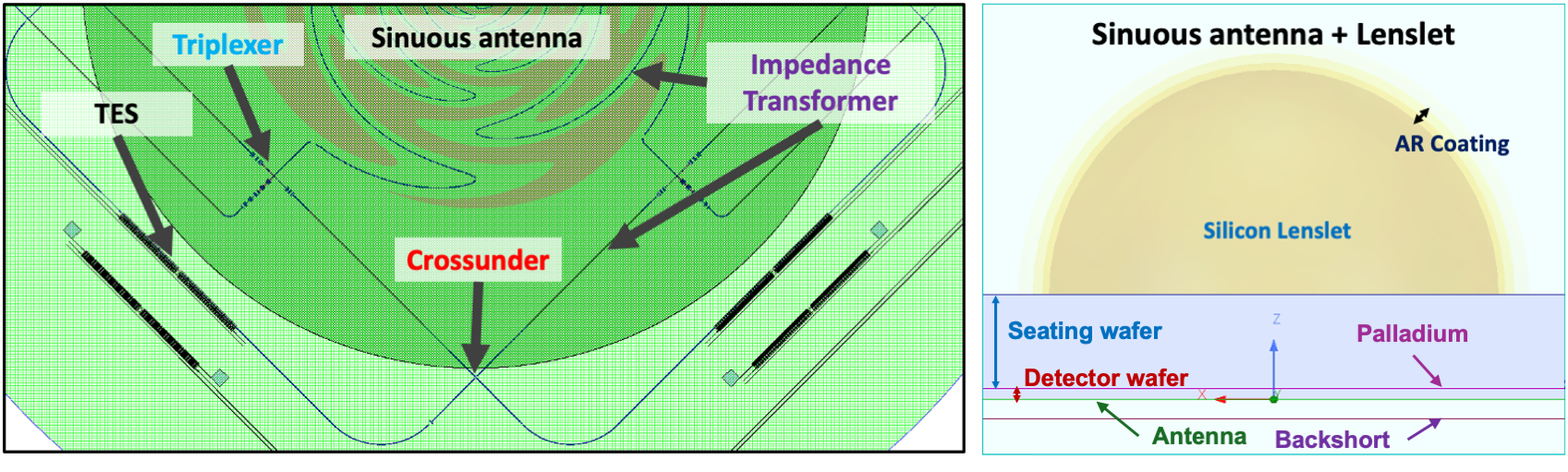}
    \caption{\textbf{Left:} A zoom-in on a LF12 wafer layout around one of the sinuous pixels. The main components of the detector chain can be seen: the sinuous antenna, the triplexer that defines the three frequency bands, impedance transformers, cross-under structures and the TES bolometers. The light-green area shows the extent of the silicon lenslet, while the dark-green area shows the negative of the palladium deposited on the sky-side of the detector wafer. \textbf{Right:} A side view of a HFSS model of the pixel structure, showing (from top to bottom) the silicon lenslet with a 3-layer AR coating, the seating wafer, the detector wafer and the metal backshort. The sinuous antenna is located on the niobium surface that crosses the origin of the reference system. The surface in purple above indicates the location of the palladium layer, while the one below in pink indicates the backshort.}%
    \label{fig:LF12_feed_pixel}
\end{figure}

In this paper we describe the design of the lowest frequency pixels of LiteBIRD Low Frequency Telescope (LFT): LF1 (40-60-78 GHz) and LF2 (50-68-89 GHz).

\section{Low Frequency Telescope Focal Plane Unit}\label{sec:lftfpu}
LiteBIRD LFT is a reflective Cross-Dragone telescope \cite{litebird_ptep} and houses a rectangular $\sim420\times 210$ mm focal plane unit (FPU) to optimize the use of the double mirror optical configuration. The focal plane unit is made of eight $\sim110\times 110$ mm focal plane sub-assemblies (FPSAs). The size of the FPSA is primarily determined by the dimensions of the detector arrays ($101\times101$ mm) which are fabricated on 6-inch silicon wafers (see Figure \ref{fig:FPUandLF12wafer}). 

As shown in Figure \ref{fig:lftfpu}, LFT spans the frequency range 34-161 GHz, from the galactic synchrotron dominated low-end of the spectrum to the main CMB science portion around 100 GHz.

LFT implements arrays of trichroic dual-polarization pixels by employing lenslet-coupled sinuous antennas. The broad band observed by the antenna is filtered using microstrip on-chip distributed filters (microfabricated on the same silicon wafer that houses the sinuous antenna and TES detectors). From each antenna we obtain three contiguous but separate 23-30\% bands cantered at the frequencies listed in Figure \ref{fig:lftfpu}. 

We implement four different pixel types (as shown in Figures \ref{fig:lftfpu} and \ref{fig:FPUandLF12wafer}): LF1 (40-60-78 GHz), LF2 (50-68-89 GHz), LF3 (68-89-119 GHz) and LF4 (78-100-140 GHz) for a total of 9 separate (but overlapping bands), with some bands present in two different pixel types (12 total frequency channels).

In Figure \ref{fig:FPUandLF12wafer} we can see that LF1 (red) and LF2 (yellow) pixels (lowest frequency range of LFT) are placed in the 4 wafers (LF12) at the edge of the focal plane unit. While LF3 (green) and LF4 (blue) are placed in the 4 wafers (LF34) at the center of the focal plane unit. 

Since in this paper we focus on the design of LF1 and LF2 pixels, in Figure \ref{fig:FPUandLF12wafer} we only report the layout of a LF12 wafer, showing a total of 9 sinuous pixels (6 LF1 and 3 LF2). In this plot we highlight the total size of the wafer and the pixel pitch (in light green), corresponding to the size (32 mm) of the silicon lenslet that is employed to boost the directivity of the antenna. LF34 wafers have the same total dimension of LF12 wafers but fit 36 pixels (18 LF3 and 18 LF4) with a pixel pitch of 16 mm.

As reported in Figure \ref{fig:lftfpu}, with the aforementioned wafer layout and number of wafers, LFT FPU will house a total of 180 dual-polarization trichroic sinuous pixels for a total of 1080 optical TES bolometers. \begin{figure}[tp]
    \centering
    \includegraphics[width=1\textwidth]{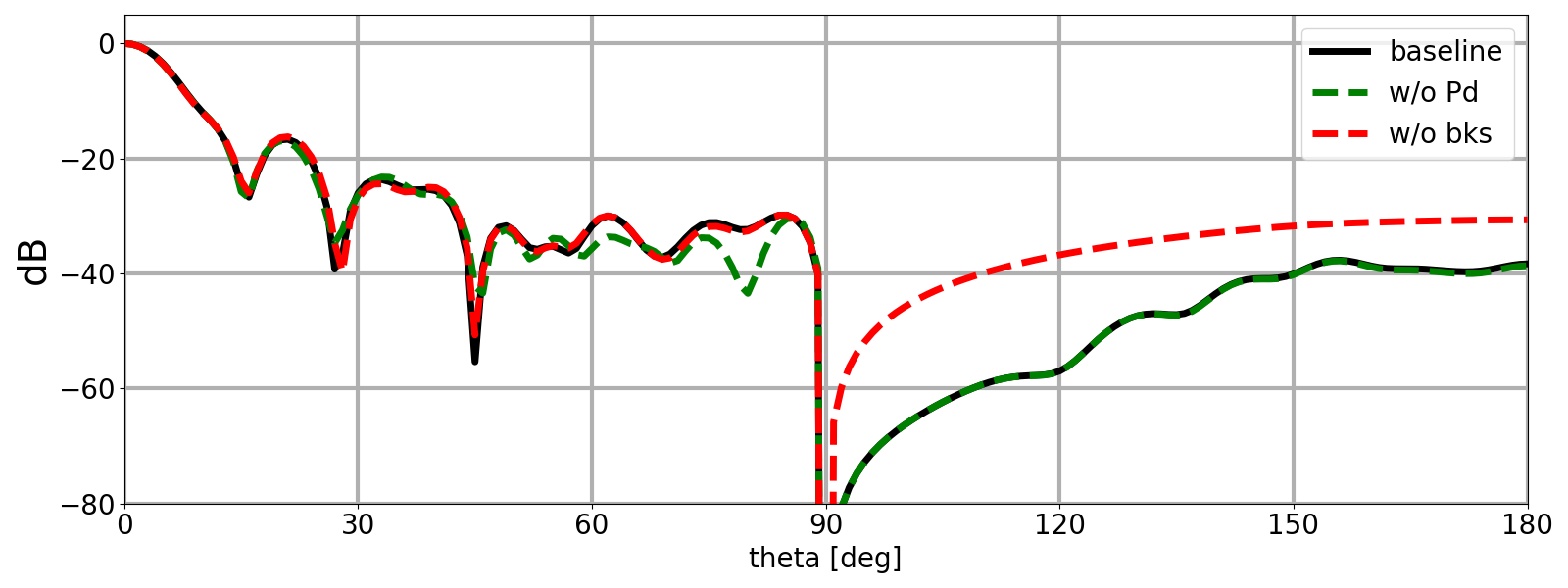}
    \caption{Comparison of the beam function at 60 GHz for $\phi = 0^{\circ}$ for the LF1 pixel for three cases: baseline (black line), baseline without palladium layer (green dashed line) and baseline without metal backshort (red dashed line). We show this to highlight that the backshort significantly reduces the back-lobe pickup as can be seen for $\theta>90^{\circ}$, and the palladium layer interfere minimally with the antenna (and only away from the main lobe as can be visibly seen in the range $60^{\circ}<\theta<90^{\circ}$) when a 20 mm diameter hole in the palladium layer is aligned with the antenna to let the radiation reach the antenna.}%
    \label{fig:beam}
\end{figure}

\section{PIXEL STRUCTURE}
\label{sec:pixel}
Lenslet-coupled sinuous antennas are quite well known in the radio astronomy community because of their dual polarization properties and broad-band nature. Because of these two properties and the race to characterize CMB polarization, the POLARBEAR/Simons Array \cite{Suzuki2016, suzuki_phd} and SPT\cite{spt3g_2022} collaborations pioneered their use for CMB observations.

After its use in Simons Array, SPT and more recently Simons Observatory, we have decide to implement the same technology to leverage the decade long experience 
developing this type of pixels and to use them in their full potential by packing the largest possible amount of detectors given the constraints in size and mass due to the nature of a space mission.

In Figure \ref{fig:LF12_feed_pixel} we show the structure of the detector chain. On the left side we highlight all the components that are part of the device wafer and are microfabricated on 6-inch silicon wafers. While on the right side we show a side view of the full pixel structure, composed of an anti-reflection (AR) coated hemispherical silicon lenslet, a silicon seating wafer to extend the lenslet and mimic an elliptical lens and enhance directivity, the detector wafer (the wafer supports the sinuous antenna and all the other components of the detector chain) and a metal backshort to suppress the a residual antenna back-lobe.

Details of each component, design choices and simulated performance are going to be explained in the following sections. 

\section{LENSLET-COUPLED SINUOUS ANTENNA}
\label{sec:antenna}
The first element of the detection chain is the lenslet-coupled sinuous antenna. All the relevant components are shown in Figure \ref{fig:LF12_feed_pixel}. The key component is the sinuous antenna which collects the radiation from the microwave sky. We tuned the antenna dimensions to maximize its efficiency in the range of interest for LF1 and LF2 pixels frequencies 34-99 GHz. After some trade-off studies we settled on an antenna with a 24 $\rm{\mu}$m inner radius, expansion factor $\tau = 1.3$ and 21 cells\footnote{$R_\mathrm{out}=R_\mathrm{in}\tau^{n_\mathrm{cell}}$. In this formula $R_\mathrm{in}$ is the radius of the innermost cell that defines the sinuous antenna, $n_\mathrm{cell}$ is the total number of cells, $\tau$ is the expansion factor defined as $R_\mathrm{i+1}/R_\mathrm{i}$ and $R_\mathrm{out}$ is the outer radius of the sinuous antenna. The inner and outer radius define respectively the highest and lowest frequency cut-offs of the antenna. While $\tau$ defines how packed are the arms of the antenna.}. With these parameters we obtain an outer radius of $\sim6$ mm. 
\begin{figure}[tp]
    \centering    \subfloat{{\includegraphics[width=.49\textwidth]{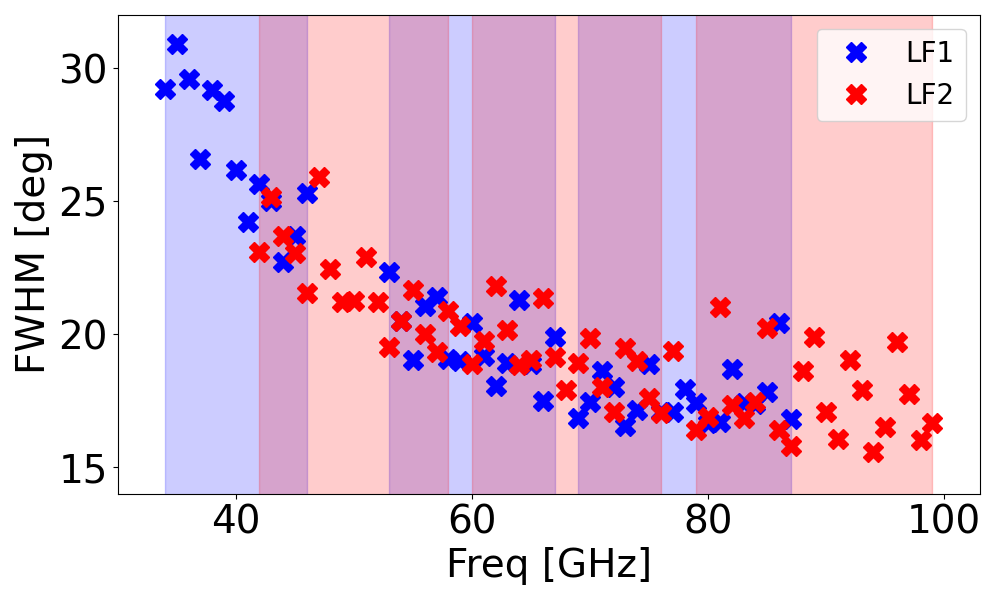} }}%
    \subfloat{{\includegraphics[width=.49\textwidth]{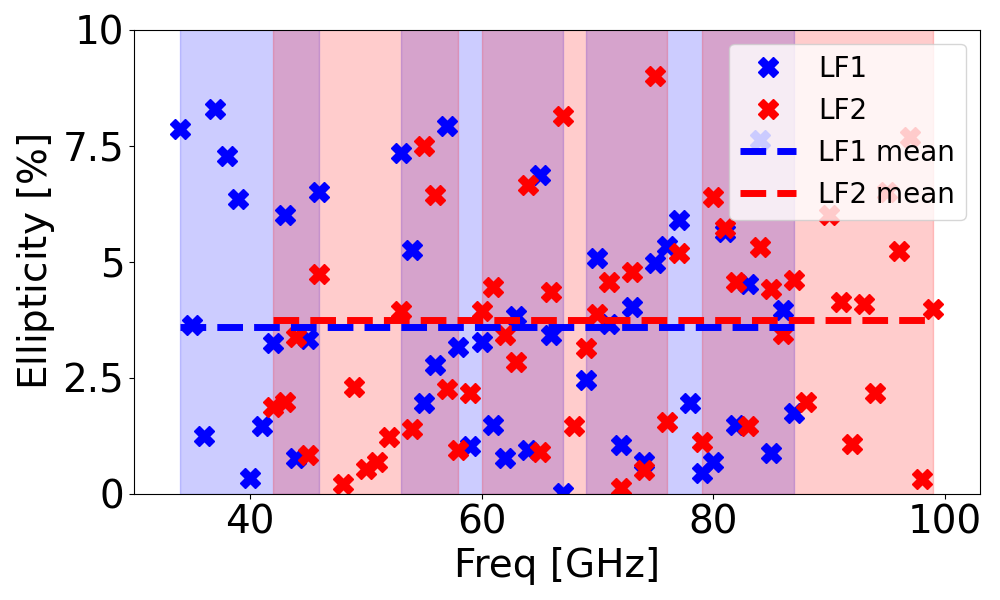} }}%
    \caption{\textbf{Left:} HFSS simulation of the frequency-dependent full-width at half-maximum of the beam function for LF1 and LF2. \textbf{Right:} Beam ellipticity and average value for LF1 and LF2. In both plots the vertical blue and red bars highlight the nominal bands of pixel LF1 and LF2 respectively.}%
    \label{fig:fwhm_ellipticity}
\end{figure}

For the trade-off study we looked predominantly at two factors: the complex impedance of the antenna in free space and the minimum dimensions of the arms of the sinuous antenna. From the theory\cite{suzuki_phd, Suzuki2012, Edwards2012} we expect that the complex impedance of a sinuous antenna in free-space should be stable in the frequency range of interest. In particular we want the impedance to be real with a value of $\sim250~\Omega$. Therefore in the first step of our optimization we simulated (HFSS\footnote{\url{https://www.ansys.com/products/electronics/ansys-hfss}}) the smallest $R_\mathrm{in}$ and $R_\mathrm{out}$ that achieve the best in-band performance in terms of complex impedance in free-space. While it is known that smaller values of $\tau$ can achieve better stability in band, the minimum dimensions of the arm of the antenna for $\tau<1.3$ become too small to support the microstrip line efficiently, in fact the width of the arms of the sinuous antenna (which define the ground plane of the microstrip structure) for $\tau<1.3$, close to the center, become comparable to the width of the microstrip line trace ($\sim 1~\mathrm{\mu m}$). In this configuration the ground plane can not be assumed to be infinite anymore and the loss of the transmission line increases.   

As discussed in \citenum{suzuki_phd} a hemispherical silicon lenslet and a silicon wafer can be used to boost the directivity of the antenna toward the sky. An elliptical lenslet would be ideal but due to difficulties in procuring and handling elliptical silicon lenslets, a stack with a hemispherical lenslet and a silicon wafer that mimics the effect of an ellipse as been used successfully in the past. From our analysis we find that a seating wafer with a thickness $L\sim0.37\times R_\mathrm{lens}$ achieves the best gain in the LF1 and LF2 frequency range and the pixel geometry described above. Given that for LF1 and LF2 $R_\mathrm{lens}\sim16$ mm, we find that this would require a very thick and difficult to machine with standard techniques silicon wafer, therefore we are currently exploring the possibility of building the extension into the lenslet itself. This solution is still being studied and evaluated by our team.

A second change compared to previous generations of this technology is the anti-reflection coating on the surface of the lenslet. An anti-reflection coating is required to minimize losses due to reflections at the boundary between free-space and silicon, due to the difference in refractive index ($n\sim3.4$ for silicon). 
A common technique that has been used in the past in the CMB community for silicon or similar materials (e.g. sapphire and alumina) to smooth the transition consists in coating the surface of the lenslet with one or more layer of epoxy. By tuning the thickness of each layer and choosing properly the refractive index of each layer it is possible to reach efficiencies well above 90\% across a broad frequency range. However an important shortcoming of this technique comes from differential thermal contractions and expansions due to the use of different materials which can cause significant problems for cryogenic applications. In order to avoid this issue, we have been developing an alternative approach where the surface of the silicon lenslet is laser-machined to ablate part of the silicon creating a sub-wavelength periodic structure\cite{beckman_phd}. Starting from the top of the lenslet the filling factor of silicon increases towards the bulk of the lenslet achieving a layer with a smooth refractive index transition from that of free-space ($n=1$) to that of silicon. This technique is not completely new in the community, however it was mostly employed on flat surfaces\cite{Takaku_21}.
\begin{figure}[tp]
    \centering
    \includegraphics[width=.98\textwidth]{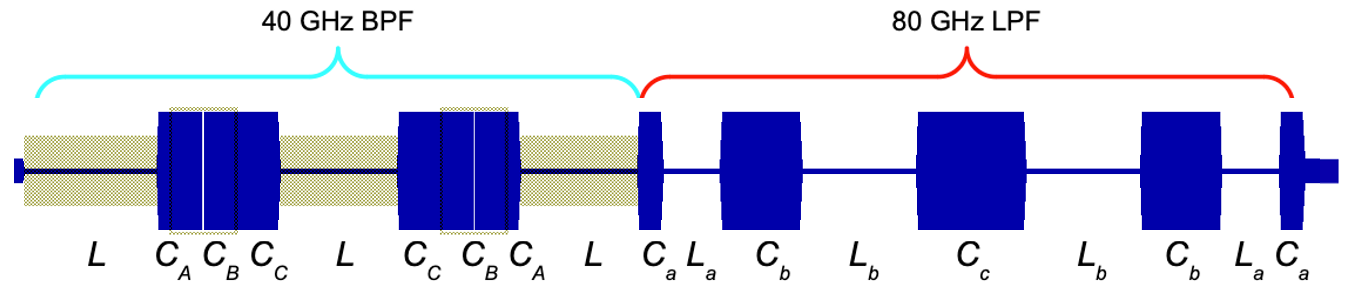}
    \caption{An example of an on-chip filter for the 40 GHz channel of LF1. Each filter is a combination of a distributed three-pole Chebyshev bandpass filter (BPF) and a distributed five-pole Chebyshev lowpass filter (LPF) to suppress high-order harmonics. The lowpass cut-off is $\sim 2\times f_c$, where $f_c$ is the central frequency of the bandpass filter.}%
    \label{fig:bpf_and_lpf}
\end{figure}

\begin{figure}[bp]
    \centering
    \includegraphics[width=1\textwidth]{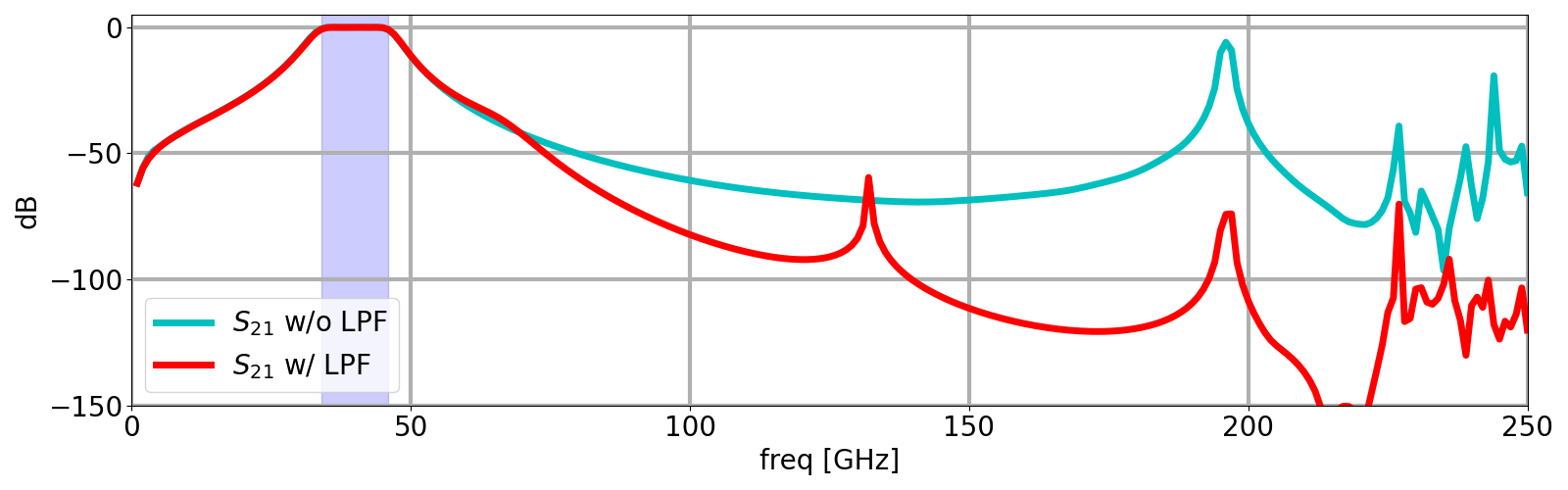}%
    \caption{Comparison of the response of the filter in Figure \ref{fig:bpf_and_lpf} in red and the bandpass filter alone in blue. We can notice a significant suppression of a high-order harmonic around 190 GHz because of the lowpass filter.}%
    \label{fig:lpf}
\end{figure}

Due to memory limitations it is very challenging to include the geometry of the anti-reflection coating in EM simulations, therefore we approximate this by calculating the effective refractive index from the AR coating structure filling factor and create a multi-layer coating in the HFSS model (see Figure \ref{fig:LF12_feed_pixel}).

There are two more differences between the LiteBIRD LFT pixel design and earlier iterations of this technology: a reflective (metal) backshort placed at $\lambda/4$ distance behind the sinuous antenna and a layer of palladium deposited on the sky-side surface of the detector wafer. The former has been introduced to mitigate the back-lobe of the antenna. This can be easily noticed in Figure \ref{fig:beam} where our baseline design in black (the geometry explained above with both the backshort and palladium layer) is compared to a design without the reflective backshort in red (dashed). We can easily notice how the back-lobe ($90^{\circ}<\theta<180^{\circ}$) is suppressed by up to a few tens of dBs in the baseline case. 

The second difference is the addition of a palladium layer on the sky-side surface of the detector wafer (in contact with the seating wafer). The reason for this change is twofold: palladium has a large heat capacity that can be leveraged to stabilize the temperature of the wafer during operation and to reduce the impact of stray light or direct absorption on the TES bolometers.
However, in order to allow the signal from the sky to reach the antenna, after deposition of the palladium layer, the palladium needs to be patterned by removing it from the optical path. After some trade-off studies we found that a circular opening of 20 mm diameter aligned with the sinuous antenna and lenslet achieve the best compromise of protecting the TES bolometer (as can be seen in Figure \ref{fig:LF12_feed_pixel}) and not-interfering with the antenna. We show this in Figure \ref{fig:beam} where the baseline design in black  is compared to a design without the palladium layer (with the 20 mm hole) in green (dashed). We can see that the main-lobe is unaffected by the presence of the palladium layer and only a small discrepancy between the two cases can be noticed in the range $60^{\circ}<\theta<90^{\circ}$.
\begin{figure}[tp]
    \centering
    \subfloat{{\includegraphics[width=.49\textwidth]{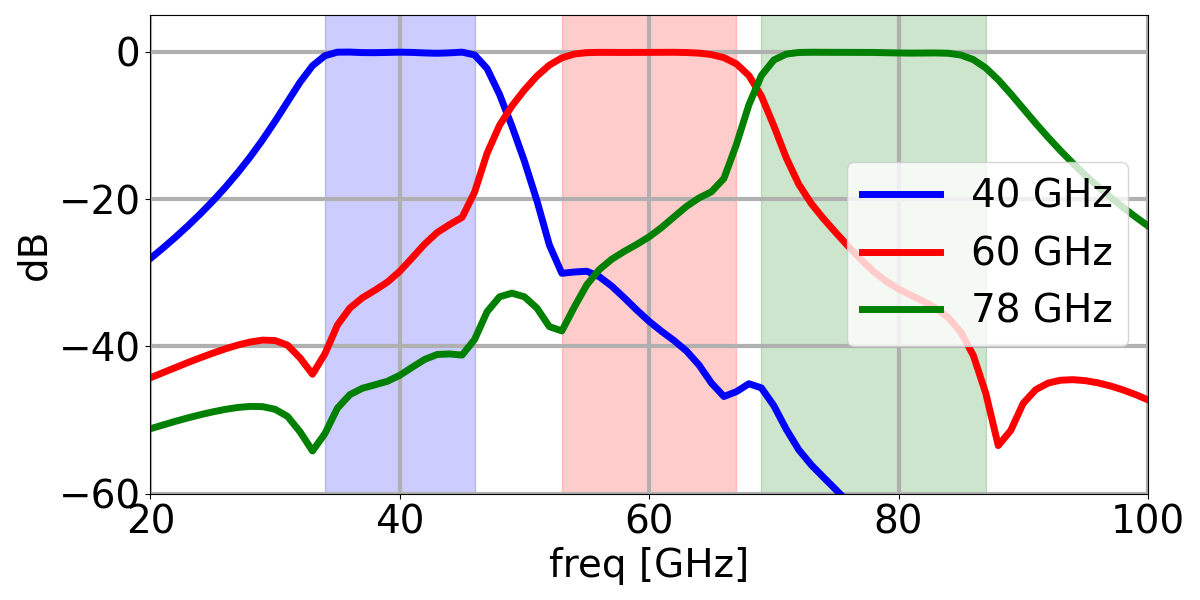} }}%
    \subfloat{{\includegraphics[width=.49\textwidth]{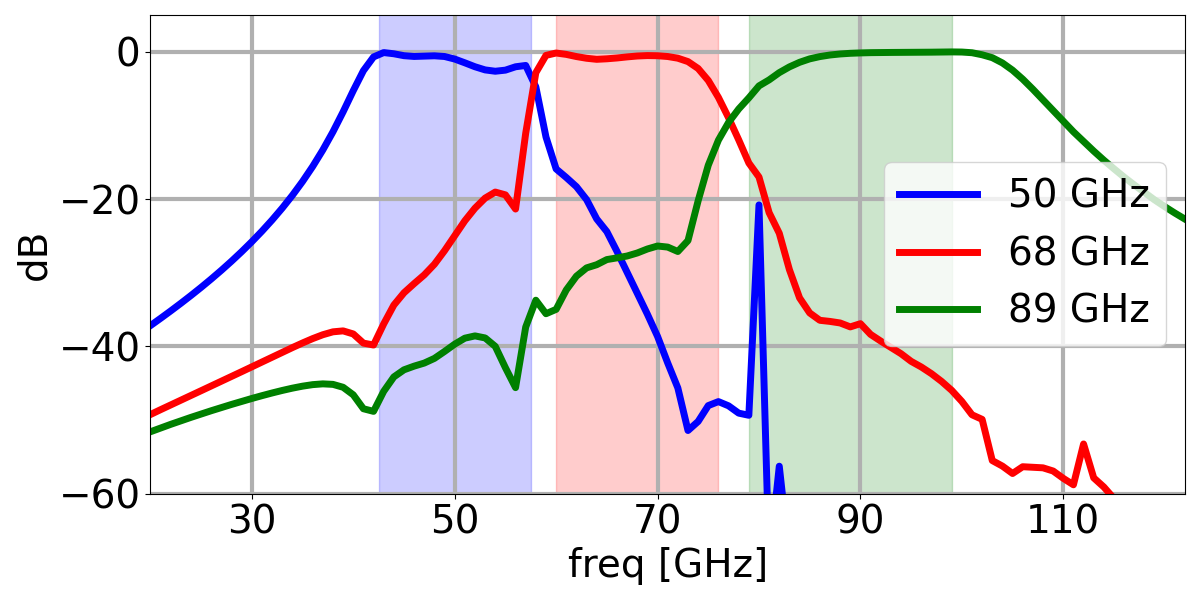} }}%
    \caption{\textbf{Left:} Simulated transmission of the LF1 pixel triplexer. \textbf{Right:} Simulated response of the LF2 pixel triplexer (one more design iteration is required to improve the response of the 89 GHz bandpass filter).}%
    \label{fig:bandpass}
\end{figure}

\begin{figure}[bp]
    \centering
    \subfloat{{\includegraphics[width=.49\textwidth]{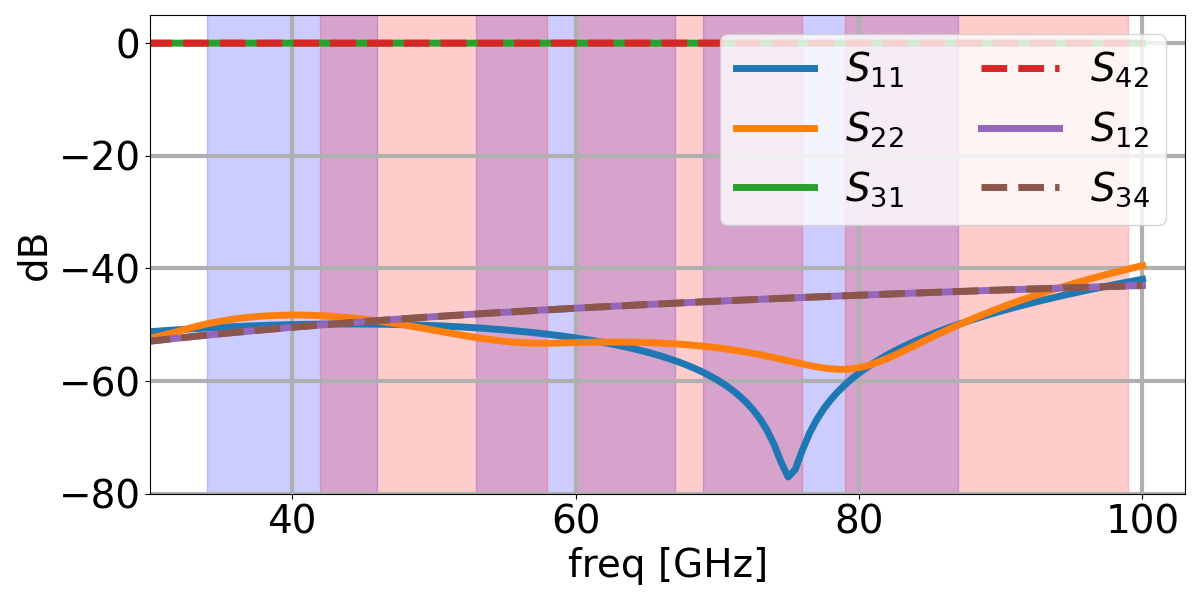} }}%
    \subfloat{{\includegraphics[width=.49\textwidth]{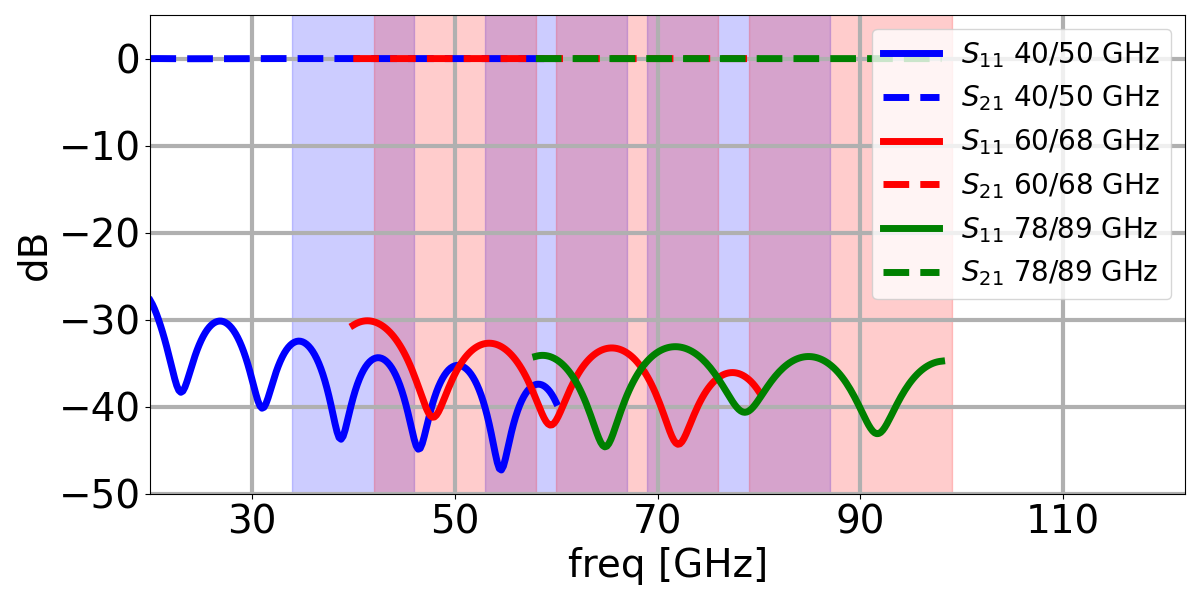} }}%
    \caption{\textbf{Left:} Simulated response of the cross-under structure showing good transmission and isolation of opposite ports in the LF1 and LF2 frequency range. \textbf{Right:} Transmission and return loss of tapered 6.8 to 12.7 $\Omega$ impedance transformers in the LF1 and LF2 frequency range. In both plots the vertical blue and red bars highlight the nominal bands of pixel LF1 and LF2 respectively.}%
    \label{fig:rfstructures}
\end{figure}

In Figure \ref{fig:fwhm_ellipticity} we show the simulated (HFSS) full-width at half-maximum (FWHM) and ellipticity for the baseline case for LF1 and LF2. 

\section{RF STRUCTURES}
\label{sec:rf_structures}
The sinuous antenna is fabricated on a niobium layer deposited on the 6-inch wafer (above a layer of silicon-nitride that is required as the structural membrane of the TES bolometer). When the broad-band signal is collected by the antenna, it needs to be filtered and routed to the TES bolometer. This step is achieved with standard microwave engineering techniques, namely by employing a microtrip line that channels the signal from the center of the antenna to the edge where TES bolometers are fabricated. 

Using microstrip line is particularly suited to this technology as the niobium layer deposited for the antenna can be used to define the ground-plane of the microstrip line. 
A layer of silicon-nitride of 1 $\mu$m thickness is deposited above the ground plane to form the dielectric of the microstrip line.

Before reaching the detectors, the signal is filtered by a triplexer that splits the broad-band signal from the antenna in narrower bands (23-30\% bandwidth). After a trade-off study of performance versus manufacturing errors we decided to implement the filters with 6.8 $\Omega$ impedance microstrip lines. This requires lowering the impedance from $\sim 50~\Omega$ at the center of the antenna to the edge using a tapered impedance transformer.

The triplexer combines three separate filters with each filter defining a band with 23-30\% bandwidth. Each filter is a combination of a distributed three-pole Chebyshev bandpass filter (BPF) and a distributed five-pole Chebyshev lowpass filter (LPF) to suppress high-order harmonics. The lowpass cut-off is $\sim 2\times f_c$, where $f_c$ is the central frequency of the bandpass filter. An example for the 40 GHz LF1 filter can be seen in Figure \ref{fig:bpf_and_lpf}. A comparison of the behaviour of the bandpass filter alone and the bandpass filter followed by the lowpass filter can be seen in Figure \ref{fig:lpf}. We can notice a significant suppression of a high-order harminic at $\sim 190$ GHz.
\begin{table}[tp]
    \centering
    \begin{tabular}{c | c c c | c c c}
        \multicolumn{1}{c}{} & \multicolumn{3}{c}{LF1} & \multicolumn{3}{c}{LF2} \\
        \hline
        Band Center [GHz] & 40 & 60 & 78 & 50 & 68 & 89 \\
        Bandwidth [\%] & 30 & 23 & 23 & 30 & 23 & 23 \\
        Optical Power [pW] & 0.2918 & 0.2419 & 0.2686 & 0.306 & 0.2709 & 0.2958 \\
    \end{tabular}
    \caption{LiteBIRD LFT LF1 and LF2 pixel design parameters for each frequency band. The fractional bandwidth (ratio between bandwidth and the central frequency) is reported. Details of the expected optical power calculation per detector can be found in \citenum{westbrook_2022} and \citenum{hasebe_2023}.}
    \label{tab:LF12}
\end{table}

\begin{figure}[bp]
    \centering
    \subfloat{{\includegraphics[width=.49\textwidth]{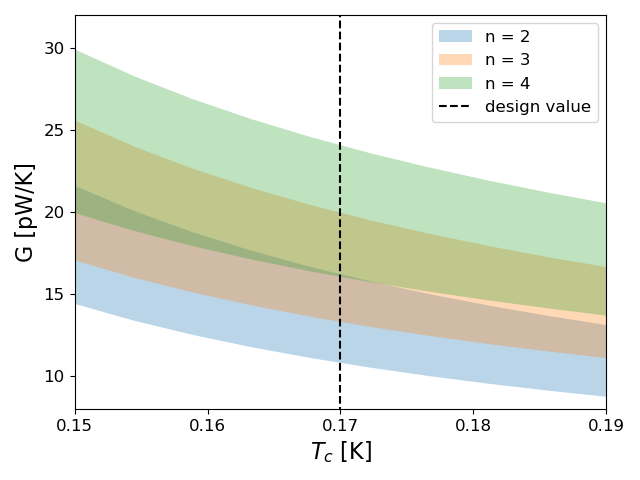} }}%
    \subfloat{{\includegraphics[width=.49\textwidth]{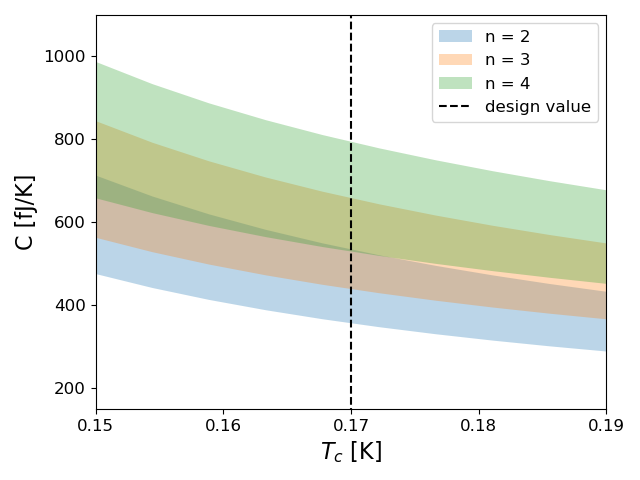} }}%
    \caption{TES bolometers thermal conductance and heat capacity target values for typical values of the thermal conductance index as a function of the transition temperature Tc of the superconductive AlMn film. The shaded area is for saturation powers $P_\mathrm{sat}$ in the range $2-3 \times P_\mathrm{opt}$. Typical value of 0.3 pW and 33 ms have been assumed for the optical power and fundamental time-constant.}%
    \label{fig:LF12_GC}
\end{figure}

The total response of the LF1 and LF2 triplexer (simulated with Sonnet\footnote{\url{https://www.sonnetsoftware.com}}) can be seen in Figure \ref{fig:bandpass}. We are aware that the response of the 89 GHz band of LF2 can be further improved in future design iterations.

Given that 6 detectors are necessary to complete the detection chain for this type of pixel, one more RF structure is necessary to avoid mixing the signal between channels: a cross-under structure. This is a four-port devices where the signal for one channel travels from port 1 to port 3 and from port 2 to port 4 for a second channel. Hence, the requirement is to maximise transmission between port 1 (2) and port 3 (4) while maintaining low return loss and good isolation between ports belonging to different channels. After some study we found better performances for a cross-under with 12.7 $\Omega$ impedance compared to the 6.8 $\Omega$ case that would match the impedance of the triplexer. We opted for the former (12.7 $\Omega$) as it also matches the impedance of the titanium load resistor on the TES island. The performance of the cross-under can be seen on the left panel of Figure \ref{fig:rfstructures}, while on the right side the performance of tapered 6.8 to 12.7 $\Omega$ impedance transformers are reported. These are necessary to minimize losses in the transition from the low impedance triplexer to the higher impedance cross-under. The impedance transformers used here are all $\sim 10$ mm long; we found that we obtained the lowest $S_{11}$ (return loss) values for a linear taper design. In Figure \ref{fig:rfstructures} we can observe that $S_{11}<-30$ dB across the LF1 and LF2 bands. Other design tested (stepped impedance and single quarter wave transformers) can achieve better performances in certain frequency ranges but not across the whole range. 

\section{TES BOLOMETERS}
\label{sec:tes}
The final component of the detection chain are the superconducting TES (transition-edge sensor) bolometers. The signal collected by the antenna after being channeled and filtered by the RF circuit described above, is dissipated on a silicon-nitride (SiN) island suspended by four thin silicon-nitride legs. The microstrip line travels along two of the legs and it is terminated by a titanium load resistor matched to the transmission line. The titanium resistor is tightly coupled to a superconductive aluminum-manganese film (AlMn) that is voltage-biased in its superconductive transition through niobium bias lines. 

The response and properties of the TES are determined by its thermal conductance G (primarily determined by the geometry of the silicon-nitride legs) and heat capacity C (a palladium thermal ballast is deposited on the TES island to tune its value).
In Table \ref{tab:LF12} we report the expected optical power loading each detector type for the LF1 and LF2 pixels. Details of this calculation and assumptions can be found in \citenum{westbrook_2022} and \citenum{hasebe_2023}. From these values and assuming a target saturation power $P_\mathrm{sat}$ in the range $2-3 \times P_\mathrm{opt}$ (optical power) and target time-constant $\tau_0\sim33$ ms, from standard bolometer equations\cite{Irwin2005} we can find the target thermal conductance and heat capacity values as a function of the TES critical temperature $T_c$ as reported in Figure \ref{fig:LF12_GC}. For the first fabrication iteration of these devices we have included multiple TES designs with different leg and thermal ballast geometries to test the thermal conductance and heat capacity dependence from design choices. This will help us narrow down the best design and incorporate changes in future fabrication iterations to quickly converge to the desired properties for LF1 and LF2 detectors. 

\section{CONCLUSIONS}\label{sec:conclusions}
In this work we have provided an update on the status of the design and development of LiteBIRD's LFT thricroic pixels LF1 (40-60-78 GHz) and  LF2 (50-68-89 GHz). We have discussed design choices motivated by LiteBIRD scientific goals. In particular we focused on the details of the optimization of the design parameters of the sinuous antenna, on-chip bandpass filters, cross-under and impedance transformers and all the RF components that define the LF1 and LF2 pixel detection chain. We have presented this work in the context of the technical challenges and physical constraints of a space mission.

While this design is rapidly maturing, compared to what has been presented in the past for these low frequency pixels, we foresee that some design-fabrication-characterization iterations are going to be needed before converging to a definitive design and fabrication process that meets the stringent requirements of a space mission. 

The first fabrication iteration for the devices presented in this work is ongoing right now (as of May 2024), this first batch of detectors will be used to test our design and the outcome of these tests will be used to update both the RF design presented in Sections \ref{sec:pixel}, \ref{sec:antenna} and \ref{sec:rf_structures} as well as the TES bolometer design discussed in Section \ref{sec:tes}.

\acknowledgments % equivalent to \section*{ACKNOWLEDGMENTS}       
 
This work is supported by the World Premier International Research Center Initiative (WPI), MEXT, Japan. TG acknowledges the support of JSPS KAKENHI Grant Number 22K14054. 

% References
\bibliography{report} % bibliography data in report.bib
\bibliographystyle{spiebib} % makes bibtex use spiebib.bst

\end{document}